\documentstyle[11pt]{article}

\newcommand{\f}{f_q}

\begin{document}
\pagestyle{plain}
\vspace{2cm}

\noindent
{\bf BORN'S PRINCIPLE, ACTION-REACTION PROBLEM\\
AND ARROW OF TIME}\\
\\

\hspace{1.3cm}
{\bf M. Abolhasani, \& M. Golshani}
\\

\hspace{1.3cm}
{\it Institute for Studies in Theoretical 

\hspace{1.3cm}
Physics and Mathematics}

\hspace{1.3cm}
{\it P.O.Box 19395-5531, Tehran, Iran}

\hspace{1.3cm}
{\it Department of Physics

\hspace{1.3cm}
Sharif University of Technology

\hspace{1.3cm}
P.O.Box 11365-9161, Tehran, Iran}

\hspace{1.3cm}
{\it ablhasan@netware2.ipm.ac.ir}

\hspace{1.3cm}
{\it golshani@ihcs.ac.ir}\\
\\

\hspace{1.3cm}
Received 27 August 1998\\
\\
\\
\noindent
We try to obtain Born's principle as a result of a subquantum heat 
death, using classical ${\cal H}$-theorem and the definition of a 
proper quantum ${\cal H}$-theorem, within the framwork of Bohm's 
theory. We shall show the possibility of solving the problem of 
action-reaction asymmetry present in Bohm's theory and the arrow of 
time problem in our procedure.
\\
\\
\noindent
{\bf 1. INTRODUCTION}\\
\\
Bohm's theory is a casual interpretation of quantum 
mechanics that was initially introduced by de Broglie[1] and then 
developed by David Bohm[2]. This theory is claimed to be equivalent to 
the standard quantum theory without having the conceptual problems of 
the latter[3]. Yet, there are some difficulties with this theory. In 
this theory the velocity of a particle is given by

\begin{eqnarray}
\dot{\bf x}=\frac{\hbar}{m} Im(\frac{\nabla\psi}{\psi})
=\frac{{\bf\nabla} S}{m},
\end{eqnarray}
where $S$ is the phase of the wave function ($\psi=R e^{i\frac{S}
{\hbar}}$). The wave function $\psi$ itself is a solution of 
Schr\"odinger equation:

\begin{eqnarray}
i\hbar\frac{\partial\psi}{\partial t}=-\frac{\hbar^2}{2m}\nabla^2 \psi
+V(\bf x)\psi.
\end{eqnarray}
By substituting $R e^{i\frac{S}{\hbar}}$ for $\psi$ in (2) , we shall 
have the following equations

\begin{eqnarray}
\frac{\partial S}{\partial t}+\frac{(\nabla S)^2}{2m}+V({\bf x})-
\frac{\hbar^2}{2m}\frac{\nabla^2 R}{R}=0.
\end{eqnarray}
\begin{eqnarray}
\frac{\partial R^2}{\partial t}+\nabla\cdot(\frac{\nabla S}{m} 
R^2)=0.
\end{eqnarray}
where (3) is the classical Hamilton-Jacobi equation with an additional 
term, $Q=-\frac{\hbar^2}{2m}\frac{\nabla^2 R}{R}$, which is called {\it 
quantum potential}. By differentiating (1) with respect to $t$ and 
making use of (3), we obtain

\begin{eqnarray}
m\ddot{\bf x}= -\nabla (Q+V),
\nonumber
\end{eqnarray}
which shows that the $\psi$-wave affects particle's motion through the 
$-\nabla Q$ term. To secure the action-reaction (AR) symmetry, we 
expect the presence of a term corresponding to particle's reaction on 
the wave, in the wave equation of motion (2). This term is not present 
there[4,5]. Some people argue that AR symmetry is a classical 
principle, and it is not necessarily applicable to quantum mechanics. 
This is not, in our view, a strong argument. Because, if one follows 
this line of thought, one might question the applicablicity of the 
concept of path (i. e. relation (1)) in Bohmian quantum mechanics. 
Then, we have to stick to the standard formulation of quantum 
mechanics.
\par
\hspace{.6cm}Another difficulty with Bohm's theory is Born's 
statistical principle. In this theory the field $\psi$ enters as a 
guiding field for the motion of particles but at same time it is 
required by the experimental facts to represent a probability density 
through $|\psi|^2$. The question is why the probability for the ensemble 
of particles has to be equal to $|\psi|^2$.
\par
\hspace{.6cm}One of foundational problems in physics is that of temporal 
asymmetry or the existence of the arrow of time. The laws of physics do 
not distinguish between the two directions of time. In spite of this, 
there is a very obvious difference between the future and past 
directions of time in our universe. This means that there is an arrow of 
time, i. e. time flows in one direction. Recently Wootters claimed that: 
time asymmetry is logically prior to quantum mechanics  and that:" the 
attempt to envision a sub-quantum theory suggests a picture of world in 
which time-asymmetric events are taken as fundamental. Unitary evolution 
would appear as a derived concept and would of course not be universal
"[6]. Our approach exemplifies Wootters's suggestion.

\hspace{.6cm}Our paper is organized as follows: After reviewing 
Valentini's quantum ${\cal H}$-theorem in section 2, we introduce our 
new quantum ${\cal H}$-theorem in section 3 and, finally, we show how 
the AR problem and the arrow of time problem are solved by our proposal.
\\
\\
\\
\noindent
{\bf 2. QUANTUM $\cal H$-THEOREM}\\
\\
Recently, A.Valentini has tried to derive the relation 
$\rho=|\psi|^2$ as the result of a statistical subquantum ${\cal H}
$-theorem[7]. He looked for a proper quantum function $f_N$ that, like 
the classical N-particle distribution function, satisfies Liouville's 
equation. He considered an ensemble of N-body systems which could be 
described by the wave function $\Psi$ and the distribution function $P$. 
Because $|\Psi|^2$ and $P$ must equalize during the assumed heat death, 
in general $P\neq|\Psi|^2$ and one can write 

\begin{eqnarray}
P(X,t)=\f(X,t)\ |\Psi(X,t)|^2,
\nonumber
\end{eqnarray}
where $\f$ measures the ratio of $P$ to $|\Psi|^2$ at the point $X$ 
($\bf x_1\cdots\bf x_N$), at time $t$. Since both $|\Psi|^2$ and 
$P$ satisfy the continuity equation ($|\Psi|^2$ due to its being a 
solution of the Schr\"odinger equation and $P$ by its very definition) 
one can easily show that $\f$ satisfies the following equation 

\begin{eqnarray}
\frac{\partial f_q}{\partial t}+\dot X\cdot\nabla f_q=0,
\nonumber
\end{eqnarray}
where $X$ denotes $\bf x_1\cdots\bf x_N$ as before. Thus, Valentini
defined his quantum $\cal H$-function in the following way

\begin{eqnarray}
{\cal H}_q=\int\ d^{3N}X\ |\Psi(X,t)|^2\ f_q(X,t)  ln f_q(X,t).
\nonumber
\end{eqnarray}
The only difference with the classical one is that $\f$ is defined in 
the configuration space while the classical $f_N$ is defined in the 
phase space and $d^{3N}X\ d^{3N}P\rightarrow |\Psi(X,t)|^2\ d^{3N}X$. 
Valentini used Ehrenfest's coarse-graining method[8]. He claims that 
{\large $\frac{d\bar {\cal H}_q}{dt}$}$\leq 0$, where $\bar{\cal H}_q$ 
is the coarse-graind ${\cal H}$-function and the equality holds in the 
equilibrium state, where $\bar \f=1$ or $\bar P=\bar{|\Psi|}^2$. Here 
$\bar P$ and $\bar{|\Psi|}^2$ are coarse grained forms of $P$ and 
$|\Psi|^2$ respectivily. Valentini termed this process a subquantumic 
heath death. Then, he showed that if a single particle is extracted 
from a large system and prepared in a state with a wavefunction $\psi$, 
its probability density $\rho$ will be equal to $|\psi|^2$, provided
that $P=|\Psi|^2$ holds for the large system. Notice that a one-body 
system not in quantum equilibrium can never relax to quantum 
equilibrium (when it is left to itself). But any one-body system 
extracted from a large system can do.
\par
\hspace{.6cm}Here, we want to modify Valentini's procedure so that one 
can directly obtain Born's principle for a one-body system. In this 
case we will be forced to use Boltzmann's procedure, and that naturally 
leads to a solution of the AR problem.\\
\\
\\
\noindent
{\bf 3. AN ALTERNATIVE QUANTUM ${\cal H}$-THEOREM}\\
\\
Consider an ensemble of one-body systems. Suppose that all 
these systems are in the $\psi$ state and that their distribution 
function is $\rho$. Furthermore, suppose that at $t=0$ we have $\rho
\neq|\psi|^2$, and define 

\begin{eqnarray}
\rho({\bf x},t)=\f({\bf x},t)\ |\psi({\bf x},t)|^2.
\end{eqnarray}
In Valentini's procedure the complexity of systems leads to heat death, 
but our systems are simple (one-body) ones. Thus, if we want to have 
heat death, we must assume that $\f$ satisfies a quantum Boltzmann 
equation 

\begin{eqnarray}
\frac{\partial f_q}{\partial t}+\dot{\bf x}\cdot\nabla f_q=J(f_q),
\end{eqnarray}
where $J(\f)$ is related to the particle reaction on its associated 
wave. Since $\rho$ satisfies a continuity equation, as the result of its 
definition, thus (5) and (6) imply that $|\psi|^2$ does not satisfy the 
continuity equation any more. In fact, the continuity equation for 
$|\psi|^2$ is changed to 

\begin{eqnarray}
\frac{\partial |\psi|^2}{\partial t}+\nabla\cdot(\frac{\nabla S}
{m}|\psi|^2)=-\frac{J(\f)}{\f} |\psi|^2.
\end{eqnarray}
This means that $\psi$ is a soluation of a nonlinear Schr\"odinger 
equation. If we want to have the quantum potential in its regular form, 
i.e. $(-\frac{\hbar^2}{2m}\frac{\nabla^2 R}{R})$, we must choose the 
nonlinear term in a particular form. The proper selection is

\begin{eqnarray}
i\hbar(\ \frac{\partial}{\partial t}+g(\f)\ )\psi=-\frac{\hbar^2}{2m}
\nabla^2 \psi,
\end{eqnarray}
where $g(\f)$ is a real function of $\f$. With the substitution $\psi=R 
e^{i\frac{S}{\hbar}}$ we shall have

\begin{eqnarray}
\frac{\partial S}{\partial t}+\frac{(\nabla S)^2}{2m}-
\frac{\hbar^2}{2m}\frac{\nabla^2 R}{R}=0.
\end{eqnarray}
\begin{eqnarray}
\frac{\partial |\psi|^2}{\partial t}+\nabla\cdot(\frac{\nabla 
S}{m}|\psi|^2)=- 2\ g(\f)\ |\psi|^2.
\end{eqnarray}
By comparing (10) with (7) we have $g(\f)=\frac{1}{2}\ \frac{J(\f)}
{\f}$. Now, we have to select $g(\f)$ in such a way that it leads to 
the equality of $\rho$ and $|\psi|^2$ (i.e. $\f=1$). Some requirements 
for such a function is:\\
\\
1- It must be invariant under $t\rightarrow -t$.\\
\\
2- It must change its sign for $\f=1$.\\

\hspace{.6cm}If we fined systems for which subquantum heat death has 
not occured[9], we shall obtain the actual form of the $g(\f)$ function. 
A proper selection is $g(\f)=\alpha(1-\f)$ where $\alpha$ is a constant. 
Then, (10) gives (with$\alpha=\frac{1}{2}$)

\begin{eqnarray}
\frac{\partial |\psi|^2}{\partial t}+\nabla\cdot(\frac{\nabla S}
{m}|\psi|^2)=(\f-1)\ |\psi|^2.
\end{eqnarray}
Now, consider the right hand side of (11) as the source of $|\psi|^2$ 
field. At any point of space where $|\psi|^2< \rho$  (i.e. $\f>$ 1), the 
source of $|\psi|^2$ is positive and, therefore, $|\psi|^2$ increases at 
that point. On the other hand, at any point of space where $|\psi|^2 > 
\rho$  (i.e.$\f<$ 1), the source of $|\psi|^2$ is negative and therefore 
$|\psi|^2$ decreases at that point. The variation of $|\psi|^2$ 
continues until $|\psi|^2$ becomes equal to $\rho$. After that, since 
both $|\psi|^2$ and $\rho$ evolve under the same velocity field ($\frac
{\nabla S}{m}$), they remain equal. To prove $\rho\rightarrow|\psi|
^2$ exactly, we introduce the following ${\cal H}_q$ function:

\begin{eqnarray}
{\cal H}_q=\int d^3x (\rho-|\psi|^2) ln (\frac{\rho}{|\psi|^2}).
\end{eqnarray}
Since $(X-Y)ln(${\Large $\frac{X}{Y}$}$)\geq 0$ for all $X,Y\geq 0$, we 
have ${\cal H}_q\geq 0$ -- the equality being relevant to the case $
\rho=|\psi|^2$. If we show that for the forgoing $J(\f)$ one has {\large 
$\frac{d{\cal H}_q}{dt}$}$\leq 0$, where again the equality is to 
relevant to $\rho=|\psi|^2$ (i.e. $\f=1$) state, we have shown that $
|\psi|^2$ becomes equal to $\rho$ finally. We write (12) in the form

\begin{eqnarray}
{\cal H}_q=\int d^3x |\psi|^2\ (\f-1) ln \f =\int d^3x\ |\psi|^2 G(\f),
\nonumber
\end{eqnarray}
where $G(\f)=(\f-1)\ ln\f $. Now, we have for {\large $\frac{d{\cal 
H}_q}{dt}$} 

\begin{eqnarray}
\frac{d{\cal H}_q}{dt}=\int d^3x  \{ \ -\nabla\cdot(\ 
\dot{\bf x}\ |\psi|^2\ G(\f)\ )
\nonumber
\end{eqnarray}
\begin{eqnarray}
-J(\f)\ \left [\ \frac{G(\f)}{\f}-\frac{\partial G(\f)}{\partial 
\f}\ \right ]\ |\psi|^2\ \},
\end{eqnarray}
where we have done an integration by parts. Passing to the limit of 
large volumes and dropping the surface term in (13) leads to

\begin{eqnarray}
\frac{d{\cal H}_q}{dt}=\int d^3x \frac{J(\f)}{\f}\{\f-1+ln\f \}\ 
|\psi|^2.
\nonumber
\end{eqnarray}
The quantity in $\{\}$, is negative for $\f<1$ and positive for $\f>1$. 
Now, since the $J(\f)$ (i.e. $\f(1-\f)$) is positive for $\f<1$ and 
negative for $\f>1$, the integrand is negative or zero for all values 
of $\f$ and we have 
\begin{eqnarray}
\frac{d{\cal H}_q}{dt}\leq 0,
\nonumber
\end{eqnarray}
where the equality holds for $\f=1$. The existance of a quantity that
decreases continuously to its minimum value at $\rho=|\psi|^2$ (i.e. 
$\f=1$) guarantees Born's principle.\\
\\
\\
\noindent
{\bf 4. THE ACTION-REACTION PROBLEM}\\
\\
In the classical gravity, matter fixes space-time geometry 
and correspondingly matter's motion is determined by the space-time 
Thus, the action-reaction symmetry is preserved. In fact, the 
nonlinearity of Einstein equations is the result of this mutual 
action-reaction. In the same way, mutual action-reaction between wave 
and particle in Bohm's theory, leads to a nonlinear Schr\"odinger 
equation. But nonlinearity does not necessarily mean particle reaction 
on the wave. Indeed nonlinear terms must contain some information about 
particle's position too. We claim that the non-linear term present in 
(8) arises from particle's reaction on its associated wave, and it 
represents information about particle's position. Considering the 
statistical nature of $\rho$, the question arises as to why particle's 
associated wave is affected by an ensemble of particles. This can be 
answered in the following ways:\\
\\
1- It is natural to expect particle's associated wave to be a function 
of particle's position, satisfing a non-linear equation of the following 
form:

\begin{eqnarray}
\pounds_{x,y,t,\Psi}\Psi(x,y,t)=0,
\nonumber
\end{eqnarray}
where $y$ represents particle's position, and the subscript $\Psi$ on
the operator $\pounds$ is an indication of the non-linearity of the 
equation with respect to $\Psi$. Consider an ensemble of particles with 
the same $\Psi$, but with different $y$, distributed according to 
$\rho$. One can look a $\psi(x,t)$, giving the average motion of the 
ensemble of particles instead of individual motions and satisfying the 
following equation:

\begin{eqnarray}
\pounds_{x,\rho,t,\psi}\psi(x,t)=0.
\nonumber
\end{eqnarray}
This equation is of type (8). The $\psi$ is not a linear conbination of 
functions $\Psi(x,y,t)$. Rather, it is an average field that gives the 
evolution of the ensemble correctly. Of course, it may not give the 
correct path of individual particles. Note, $\rho$ in (8) is the 
reaction of the ensemble of particles on the wave that represents 
ensemble's evolution. After quantum heat death and the equality of 
$\rho$ with $|\psi|^2$, this equation takes the form of the 
Schr\"odinger equation:

\begin{eqnarray}
\pounds_{x,t}\psi(x,t)=0.
\nonumber
\end{eqnarray}
This means that the AR problem is a result of the establishment of 
Born's principle. In fact this is similar to the argument that 
Valentini presents[10] to show that the signal-locality (i.e. the 
absence of practical instantaneous signalling) and the uncertainty 
principle are valid if and only if $\rho=|\psi|^2$.\\
\\
2- The information of a wave about particles position could be 
incomplete and this defect could be a result of the form of the 
interaction between the two. Thus, $\rho$ does not represent a 
statistical distribution for an ensemble of particles. Rather, it 
represents the maximum amount of informatiom of particle's associated 
wave about particle. Then, if we have an ensemble of particles having 
a same $\psi$ and $\rho$ we can take $\rho$ as the distribution function
for the ensemble.\\
\\
\\
\noindent
{\bf 5. ARROW OF TIME PROBLEM}\\
\\
The Arrow of time problem can essentially be seen as 
arising from the different behavior of the time asymmetric evolution 
of macroscopic irreversible processes and the time symmetric evolution 
of the reversible processes governing the underlying dynamics of the 
atomic constituents. The apparent paradox is a direct consequence of 
the symmetric nature of the basic laws of physics with respect to time 
inversion.
\hspace{.6cm}The nonlinear Schr\"odinger equation that we introduced is 
not time reversal, although after the subquantum heat death with the 
vanishing of the $g(\f)$ term, it seems time reversal. In other words, 
the equation (8) for one of the two directions of time  leads to the 
common Schr\"odinger equation that describes real microscopic world, 
but for the other direction of time, the nonlinear term $g(\f)$ grows 
up and (8) could not describe the real world. Now, if $\psi(x,t)$ is a 
solution of the common Schr\"odinger equation, so is $\psi^*(x,-t)$, 
and both $\psi(x,t)$ and $\psi^*(x,-t)$ are solutions of our nonlinear 
equation in the subquantum equilibrium state. But in the framework of 
our theory $\psi(x,t)$ and $\psi^*(x,-t)$ are not time reversal of each
other. Nevertheless, we can formally relate them to each other with a 
time reversal transformation (ignoring the nonlinear term in subquantum 
equilibrium state).\\
\\
\\
\noindent
{\bf 6. CONCLUSIONS}\\
\\
We have shown that the Born's principle can be the result 
of the presence of a 
nonlinear term in Schr\"odinger equation, with special characteristics. 
Then, we have shown that this nonlinear term can be considered as 
indicator of particle's reaction on the wave. Besides, our nonlinear 
equation introduces an arrow of time.\\
\\
\\
\noindent
{\bf ACKNOWLEDGMENTS}\\
\\
The authors would like to thank Prof. J. T. Cushing for 
useful comments on an earlier draft of this paper.\\
\\
\\
\noindent
{\bf REFERENCE}\\
\\
1. L. de Broglie, {\it J.Phys}. (Paris) {\bf 6}, 225 (1927).\\
2. D. Bohm, {\it Phys. Rev}. {\bf 85}, 166-179, 180-193 (1952).\\
3. P. Holland. {\it The Quantum Theory of Motion} (Cambridge 
University Press, Cambridge, 1993).\\
4. J. Anandan and H. R. Brown, {\it Found. Phys}. {\bf 25}, 349 
(1995).\\
5. E. Squires, {\it The Mystery of Quantum World} (Institute 
of Physics Publishing, London, 1994).\\
6. W. K. Wootters, "Is Time Asymmetry Logically Prior to 
Quantum Mechanics?", in {\it Physical Origins of Time}, edited by J. J. 
Halliwell, J. P'erez-Mercader and W. H. Zurek, (Cambridge University 
Press, Cambridge, 1994).\\
7. A. Valentini, {\it Phys. Lett}. {\bf A156}, 5 (1991).\\
8. R. L. Liboff. {\it Kinetic Theory}. (Prentice-Hall, 
New Jersey, 1990); H. J. Kreuzer. {\it Nonequilibrium Thermodinamics 
and its Statistical Foundations} (Oxford University Press, New York, 
1984).\\
9. A. Valentini, {\it On the pilot-wave theory of classical, 
quantum and subquantum physics} (Springer-Verlag, Berlin, 1996).\\
10. A. Valentini, {\it Phys. Lett}. {\bf A158}, 1 (1991).

\end{document}